\documentclass[twocolumn,aps,preprintnumbers,amsmath,amssymb,showpacs,superscriptaddress,showpacs,nofootinbib]{revtex4}
\usepackage{graphicx}
\usepackage{subfigure}
\usepackage{amsmath, amssymb}
\usepackage{multirow}
\usepackage{hhline}

\usepackage{times}

\newcommand{\degree}{^\circ}
\def \<{\langle}
\def \>{\rangle}

\begin{document}
\title{Cosmology with Coupled Gravity and Dark Energy}

\author{Ti-Pei Li}
\affiliation{Department of Physics \& Center for Astrophysics, Tsinghua University, Beijing 100084, China}
\affiliation{Key Lab. of Part. Astrophys., Inst. of High Energy Phys., Chin. Acad. of Sci., Beijing 100049, China}

\date{\today}

%\preprint{IPMU14-0012}

\begin{abstract}
Dark energy is a fundamental constituent of our universe, its status in the cosmological field equation should be equivalent to that of matter gravity. Here we construct a dark energy and matter gravity coupling (DEMC) model of cosmology in a way that dark energy  and matter are introduced into the cosmological field equation in parallel with each other from the beginning. The DEMC  universe possesses a composite symmetry from global Galileo invariance and local Lorentz invariance. The observed evolution of the universe expansion rate at redshift $z>1$ is in tension with the standard LCDM model, but can be  well predicted by the DEMC model from measurements of only nearby epochs.  The so far most precise measured expansion rate at high $z$ is quite a bit slower than the expectations from LCDM, but remarkably consistent with that from DEMC. It is hoped that the DEMC scenario can also help to solve other existing challenges to cosmology: large scale anomalies in CMB maps and large structures up to $\sim 10^3$ Mpc of a quasar group.  The DEMC universe is a well defined mechanical system. From  measurements we can quantitatively evaluate its total rest energy, present absolute radius and  expanding speed.\\

\end{abstract}

\maketitle

\section{Introduction}\label{sec:intr}
  The field equation of general relativity (GR) proposed by Einstein \cite{ein16} in 1915 is
\begin{equation}\label{ee0}
 G_{\mu\nu}=-8\pi G T_{\mu\nu}
\end{equation}
 with $G_{\mu\nu}$ being the Einstein tensor, $G$ the Newton's gravitational constant, $T_{\mu\nu}$ the energy-momentum tensor, where the natural units $c=\hbar=1$ and the sign convention (+,-,-,-) for both metric and Einstein tensor are used. In order to construct a spatially finite static universe, Einstein \cite{ein17} in 1917 added a cosmic term with a positive cosmological constant $\lambda$ ($\lambda>0$) to his primary field equation to modify it into
\begin{equation}\label{eez}
G_{\mu\nu}+\lambda g_{_{\mu\nu}}=-8\pi G T_{\mu\nu}
\end{equation}
where $g_{_{\mu\nu}}$ is the metric of spacetime.  After Hubble's discovery of the cosmic expansion, Einstein withdrew the $\lambda$-term in Eq.\,(\ref{eez}).

   It is widely believed that Einstein introduced the cosmological constant into his field equation as a repulsive energy that fills "empty" space to prevent the universe from collapsing by the gravitational pull of the matter (see, e.g.\,\cite{pea} -- \cite{lan06}). Consequently, when the cosmic expansion is observed to be accelerating \cite{rie98,per99}, which suggests the presence of universal repulsive dark energy, Eq.\,(\ref{eez}) has been used again as a  fundamental equation of the standard cosmological model LCDM.

   However, the positive cosmological constant $\lambda$ introduced by Einstein represents in fact just "{\sl the actual mean density of the matter in the universe}" \cite{ein17}. In his original paper Einstein stated:

\leftskip 7mm \rightskip 8mm \noindent  "{\sl the newly introduced universal constant $\lambda$ defines both the mean density of distribution $\rho$ which can remain in equilibrium and also the radius $R$ and the volume $2\pi^2 R^3$ of spherical space.}"

\leftskip 0mm \rightskip 0mm
\noindent In a letter to Willem de Sitter on 14 June 1917 Einstein \cite{ein17a} explained it more clearly:

\leftskip 7mm \rightskip 8mm \noindent  "{\sl In my paper, the density $\rho$ is a matter density when the matter assembled in stars is uniformly distributed in the interstellar space.}"

\leftskip 0mm \rightskip 0mm

 Therefore, in Einstein's field equation with a positive cosmological constant in LCDM, the repulsive dark energy, as another fundamental composition of the universe besides matter, has not been included yet, and how introducing dark energy into the cosmological field equation needs to be re-considered.

\section{Cosmological Equations}\label{sec:equation}
\subsection{Field Equation}
 The Newtonian equation for gravitational field $\phi_{_m}$ with density $\rho_{_m}$ of matter is
 \begin{equation}\label{ee0a}
  \nabla^2\phi_{_m}=4\pi G \rho_{_m}\,.  \end{equation}
 The Poisson equation (\ref{ee0a}) in the framework of Galilean spacetime is a starting point to construct Einstein's field equation of GR and a standard to test the correctness of the constructed GR field equation. To construct a cosmological model where the dark energy is another fundamental constituent besides matter, we should also firstly write out the field equation for dark energy in Galilean framework, rather than just simply add a constant term to the gravitational field equation (\ref{ee0}) of GR, which was established in 1916 when matter's gravity was the only known force effective at cosmological scales.

 It is natural that the field $\phi_{_\lambda}$ of repulsive dark energy  coupled with matter has the same form with that of matter gravity but opposite sign, i.e. the field equation of dark energy with density $\rho_{_\lambda}$ in Galilean framework should be
  \begin{equation}\label{el0}
  \nabla^2\phi_{_\lambda}=-4\pi G \rho_{_\lambda}\,.  \end{equation}
  From Eq.\,(\ref{el0}), the dark energy with density $\rho_{_\lambda}$ can be regarded as a kind of matter, it has also positive inertial mass with density $\rho_{_\lambda}$, but negative gravitational mass with density $-\rho_{_\lambda}$ from comparing with the field equation (\ref{ee0a}) of matter with density $\rho_{_m}$.

For cosmological consideration, matter is smoothed to be uniformly distributed with a mean density $\bar{\rho}_{_m}$, the cosmological field equation for a universe constituted  from both matter and dark energy in Galilean framework should be
\begin{equation}\label{fe}
  \nabla^2\phi=4\pi G (\bar{\rho}_{_m}-\rho_{_\lambda})\,.  \end{equation}

For a physical world constituted by matter with density $\rho_{_m}$  and non-negligible dark energy with $\rho_{_\lambda}$, the "inertial mass"  is  evaluated in terms of  $\rho_{_m}+\rho_{_\lambda}$, but the "gravitational mass" should be evaluated in terms of $\rho_{_m}-\rho_{_\lambda}$, where the "inertial mass" is not equal to the "gravitational mass". Consequently, the equivalence principle is not valid yet for cosmology with both gravity and dark energy, and the general theory of relativity is no longer  a suitable framework for cosmology as well, we thus have to work with the Galilean framework from the start.

 The cosmological field equation (\ref{fe}) describes a dark energy and matter gravity coupling (DEMC) universe driven by the resultant force of attractive gravity and repulsive dark energy. In contrast,  the field equation in Galilean framework corresponded to Einstein's field equation (\ref{eez}) is
\begin{equation}\label{fen} \nabla^2\phi=4\pi G(\bar{\rho}_{_m}\, +\rho_{_\lambda}) \end{equation}
with $\rho_{_\lambda}=\lambda/8\pi G$, it shows that the Einstein's cosmological field equation includes only gravity, not any repulsive energy. If the density $\rho_{_\lambda}$ really describes a repulsive energy, then, not the "inertial density", but the "gravitational density" $\bar{\rho}_{_m}\, -\rho_{_\lambda}$ should be included in the right hand side of the Poisson Eq.\,(\ref{fen}).  It is obvious that Eq.\,(\ref{fen}), the Newton's limit of Einstein cosmological equation, is not a proper cosmological equation in Galilean framework.

  The Neumann-Seeliger paradox for Newton's theory states that a cosmological model for uniformly distributed matter cannot be constructed from the gravitational field,  because from Newton's field equation (\ref{ee0a})
 the field intensity  $ |-\nabla\phi_{_m}\, | \sim \rho_{_m} R \rightarrow \infty$.
 However, for a DEMC universe, it is governed by the universal density difference, $\bar{\rho}_{_m}-\rho_{_\lambda}$, of two coupled fundamental cosmic fields. The Neumann-Seeliger paradox can be avoided as the combined field of mater and dark energy vanishes  with $\bar{\rho}_{_m}=\rho_{_\lambda}$ for the Universe and  inertial frames of reference can exist globally. Consequently, it is able to construct a cosmological model from the field equation (\ref{fe}) with the Galilean framework.

\subsection{Dynamic Equations}
\subsubsection{Energy Equation}
 The total rest energy of the universe with the field Eq.\,(\ref{fe}) in the framework of Galilean spacetime
\begin{equation}\label{trest} E_{rest}=\rho V=(\bar{\rho}_{_m}+\rho_{_\lambda}) V=\mbox{constant}\,, \end{equation}
where  the universe volume $V=4\pi R^3/3$ with $R$ being the universe radius. The energy density
\begin{equation}\label{rest} \rho=\bar{\rho}_{_m}+\rho_{_\lambda}=\rho_{_0} a^{-3}\,, \end{equation}
where $\rho_{_0}$ is the current energy density, the scale factor $a=R/R_0$, and $R_0$ the current radius.

 For an expanding universe, the kinetic energy
 \begin{equation}\label{ke}
 E_{k}=\frac{1}{2}E_{rest}\dot{R}^2=\frac{2\pi\rho_0 R_0^5}{3} \dot{a}^2\,, \end{equation}
 and the potential energy
 \begin{eqnarray}\label{pe}
 E_{p}&=&-\frac{4\pi G}{3}E_{rest}(\bar{\rho}_{_m}-\rho_{_\lambda})R^2 \nonumber\\
 &=& -\frac{16\pi^2G R_0^5\rho_0}{9}(\bar{\rho}_{_m}-\rho_{_\lambda})a^2\,. \end{eqnarray}
  From the conservation of mechanical energy
  \[ E_{mech}=E_{k}+E_{p}=\mbox{constant}\,, \]
   it is easy to derive  the following energy equation in DEMC cosmology
\begin{equation}\label{f1}
\dot{a}^2=\frac{8\pi G}{3}(\bar{\rho}_{_m}-\rho_{_\lambda}) a^2+\epsilon\,,
\end{equation}
where the constant
\begin{equation}\label{epsilon} \epsilon\equiv 3E_{mech}/(2\pi \rho_0 R_0^5)=2\xi/R_0^2 \end{equation}
with the ratio of mechanical to rest energies
\begin{equation}\label{xidef} \xi\equiv E_{mech}/E_{rest}\,. \end{equation}

In LCDM cosmology, Friedmann's energy equation is derived \cite{fri22} from the time-time component of Einstein's field equation (\ref{eez}) as
  \begin{equation}\label{fri}
 \dot{a}^2 =  \frac{8\pi G}{3}(\bar{\rho}_{_m}+\rho_{_\lambda}) a^2+\epsilon\,,
\end{equation}
where the mean matter density $\bar{\rho}_{_m}=\bar{\rho}_{_{m,0}}a^{-3}$ with $\bar{\rho}_{_{m,0}}$ being the present mean density of matter, and $\epsilon$ a constant.

In DEMC cosmology, the density difference $\bar{\rho}_{_m}-\rho_{_\lambda}$ in the energy Eq.\,(\ref{f1}) comes from the potential energy $E_p$, where $\bar{\rho}_{_m}-\rho_{_\lambda}$ in  Eq.\,(\ref{pe}) is the "gravitational density" of the combined field $\phi=\phi_{_m}+\phi_{_\lambda}$. Whereas the summation $\bar{\rho}_{_m}+\rho_{_\lambda}$ in Eqs.\,(\ref{trest}) and (\ref{rest}) of rest energy is the "inertial density" of the combined cosmic field.   The "inertial density" has to be separated from the "gravitational density"  when $\rho_{_\lambda}$ cannot be ignored. Therefore, Friedmann's equation (\ref{fri}) with $\bar{\rho}_{_m}+\rho_{_\lambda}$ in the term of the potential energy is obviously not proper for an universe constituted by matter and dark energy. Confusing the two kinds of energy density  could be a reason why such an impropriety in LCDM cosmology has not been recognized  for a long time.

It is interesting to see why  Friedmann's equation (\ref{fri}) with only pulling energy can interpret the acceleration of the universe. With a constant $\rho_{_\lambda}$ and after dark energy domination, the increase of the scale factor $a$ with expansion causes the first term at the right-hand side of Eq.\,(\ref{fri}) increasing. Then, from Eq.\,(\ref{fri}) with a constant $\epsilon$ (conservation of mechanical energy), the expansion rate $\dot{a}$ at the left-hand site of Eq.\,(\ref{fri}) has to be increasing as well. Such an interpretation implies a fantastic physics: the total mass of the universe is infinitely increasing with time and the added mass is immediately and completely converted into the kinetic energy of expansion. However, applying Friedmann's  Eq.\,(\ref{fri}) with a constant $\rho_{_\lambda}$ to an expanding universe is just to use a conservation of energy equation to evaluate a process where the energy is supposed to be not conserved, will inevitably cause misconceptions in physics:  that the cosmic expansion can be accelerated by a gravitational force $(\lambda>0)$ or decelerated by a repulsive force $(\lambda<0)$ from Eq.\,(\ref{fri})  results from violating the conservation law of energy in LCDM.

\subsubsection{Equation of Motion}
 For a dark energy and gravity coupled expanding universe described by the field Eq.\,(\ref{fe}) with radius $R$, Newton's second law of motion reads
 \begin{equation}\label{n2} \ddot{R}=F/E_{rest}\,. \end{equation}
 Substituting the rest energy ("inertial mass")
 \[ E_{rest}=(\bar{\rho}_{_m}+\rho_{_\lambda})V \]
 and the driving force from the "gravitational mass"
\[ F= -G\frac{\bar{\rho}_{_m}V}{R^2}+G\frac{\rho_{_\lambda}V}{R^2}=-\frac{4\pi G}{R_0\rho_{_0}}(\bar{\rho}_{_m}-\rho_{_\lambda})a^3 \]
into Eq.\,(\ref{n2}), we get the following cosmological equation of motion
\begin{equation}\label{f2}
\ddot{a}=-\frac{4\pi G}{\rho_0 R_0}(\bar{\rho}_{_m}-\rho_{_\lambda}) a^3\,.
\end{equation}
The above equation shows that the universal expansion is driven by the energy difference $\rho_{_d}=\bar{\rho}_{_m}-\rho_{_\lambda}$: for a phase of matter domination ($\rho_{_d}>0$), dark-energy domination ($\rho_{_d}<0$), or equilibrium ($\rho_{_d}=0$), the universe is decelerated, accelerated, or uniformly expanding, respectively.

In LCDM cosmology, the following  approach is widely used in textbooks on cosmology, e.g. \cite{pea,ryd03}, to deduce the cosmological equation of motion. Applying  the first law of thermodynamics to a comoving volume of an adiabatic expanding universe with energy density $\rho$ and pressure $P$ gives the fluid equation
 \begin{equation}\label{law1}
 \dot{\rho}\,\frac{a}{\dot{a}}=-3(\rho+P)\,.
 \end{equation}
 Taking the time derivative of the first Friedmann's Eq.\,(\ref{fri}) yields
 \begin{equation}\label{dfri}
  \frac{\ddot{a}}{a}=\frac{4\pi G}{3}(\dot{\rho}_{_m}\frac{a}{\dot{a}}+2\rho_{_m}+\dot{\rho}_{_\lambda}\frac{a}{\dot{a}}+2\rho_{_\lambda})\,. \end{equation}
 With assumption that the fluid Eq.\,(\ref{law1}) is valid for matter  and dark energy separately, then from Eq.\,(\ref{dfri})  one could get the second Friedmann's equation
\begin{equation}\label{fri2}
 \ddot{a}=-\frac{4\pi G}{3}(\bar{\rho}_{_m}+3P_m+2P_{_\lambda})a\,, \end{equation}
  where the pressures of matter
 \begin{equation}\label{wm}
 P_{_m}=\omega_{_m} \rho_{_m} \end{equation}
 with $\omega_{_m}$ being the parameter  of equation of state, and the pressure of dark energy
 \begin{equation}\label{wl}
 P_{_\lambda}=-\rho_{_\lambda}\,. \end{equation}

 Eq.\,(\ref{wl}) is derived from Eq.\,(\ref{law1}) for the dark energy with $\dot{\rho}_{_\lambda}=0$.
 The fluid Eq.\,(\ref{law1}) is another manifestation of energy conservation, however,  the dark energy is not conserved for an expanding volume with a constant $\rho_{_\lambda}$, and then the fluid Eq.\,(\ref{law1}) cannot be applied to it at all. That the universal field with $\rho_{_\lambda}>0$ has a repulsive pressure as shown by Eq.\,(\ref{wl}) results again from violating the conservation of energy.
By making use of Eq.\,(\ref{wl}), the standard model makes the positive constant $\rho_{_\lambda}$ to produce a repulsive pressure. However, from the  first Friedmann's equation (\ref{fri}),
  the same constant  $\rho_{_\lambda}$  behaves just like another gravitational pull source addition to the matter. This contradiction indicates that interpreting the cosmic expansion by means of equations of state or pressures is not a correct approach.

Cosmologists often use an expanding blown balloon  to illustrate the expansion of the universe: dots  on the surface of the balloon  representing galaxies move apart from each other but stay the same size.   It is obvious that we have to separate  the field source that drives the global expansion of the universe (the blowing of pump or mouth in the balloon analogy) from that governs the local motions within a galaxy (within a dot in the balloon analogy).  The motion equation in DEMC cosmology,  Eq.\,(\ref{f2}), can meet the requirement: the expanding universe is indeed driven by the universal gravity and dark energy. On the contrary, Eq.\,(\ref{fri2}) makes use of the pressure of matter.  The pressure $P_{_m}$  of matter, including radiation pressure, is a physical quantity to describe a local system, therefore, in LCDM cosmology the global expansion of the universe is governed by local processes of matter (just like the balloon's expansion being governed by physical processes within dots). At cosmological scales, all local inhomogeneities of matter specified by $T_{\mu\nu}$ are smoothed out to be homogeneous with a density of $\bar{\rho}_m$. The global expansion of the universe should be driven by the intrinsic gravitational pull of $\bar{\rho}_m$, noting to do with the state of matter. The universe expansion causes the temperature to decrease, then the state of matter to change, but not vice versa.

\subsubsection{Expansion Equation}
In DEMC cosmology, the density $\bar{\rho}_{_m}-\rho_{_\lambda}$  appears in the energy equation (\ref{f1}) in the term representing the potential energy of the combined cosmic field, and also  in the motion equation (\ref{f2}) representing  the force driving the expansion as well.  Eliminating it by combining equations (\ref{f1}) and (\ref{f2}), we get a simple expansion equation linking the universe scale factor $a$, expansion rate $\dot{a}$ and acceleration $\ddot{a}$
\begin{equation}\label{evo}
\ddot{a}+\mu(\dot{a}^2-\epsilon) a=0
\end{equation}
where the constant $\mu$ is defined by
\begin{equation}\label{mu} \mu\equiv\frac{3}{2\rho_0 R_0}\,. \end{equation}

The expansion equation (\ref{evo}) has three solutions \cite{lw13}:
\begin{widetext}
\begin{eqnarray}
\dot{a} & =& \sqrt{\epsilon-c_1\exp(-\mu a^2)} \hspace{9mm} \mbox{for}~\dot{a}<\dot{a}_c \hspace{6mm} (\ddot{a}> 0,~ \mbox{acceleration}) \label{s1}\\
\dot{a} & =& \sqrt{\epsilon} \hspace{34mm} \mbox{for}~\dot{a}=\dot{a}_c \hspace{6mm} (\ddot{a}=0, ~   \mbox{constant expansion}) \label{s2} \\
\dot{a} & =& \sqrt{c_2\exp(-\mu a^2)+\epsilon} \hspace{9mm} \mbox{for}~\dot{a}>\dot{a}_c \hspace{6mm} (\ddot{a}< 0,~ \mbox{deceleration}) \label{s3}
\end{eqnarray}
\end{widetext}
where $c_1$ and $c_2$ are positive integral constants, and the critical expansion rate
\begin{equation} \dot{a}_c=\sqrt{\epsilon}=\sqrt{2\xi}/R_0. \end{equation}

\section{Expansion Rate}\label{sec:rate}
The solutions (\ref{s1})-(\ref{s3})  present a specific evolution picture: the  universe may remain in an equilibrium state [solution (\ref{s2})] at a constant expansion rate of $\dot{a}_c=\sqrt{\epsilon}$
 with balanced gravity-dark energy $\bar{\rho}_{_m}=\rho_{_\lambda}$ and acceleration $\ddot{a}=0$. It occasionally deviates by a phase transition with conversion between matter and dark energy, then relaxes back to the equilibrium state again along with expansion through a temporal acceleration or deceleration phase [solution (\ref{s1}) or (\ref{s3})].

\begin{figure}
   \begin{center}
\includegraphics[height=6.5cm, width=8cm, angle=0]{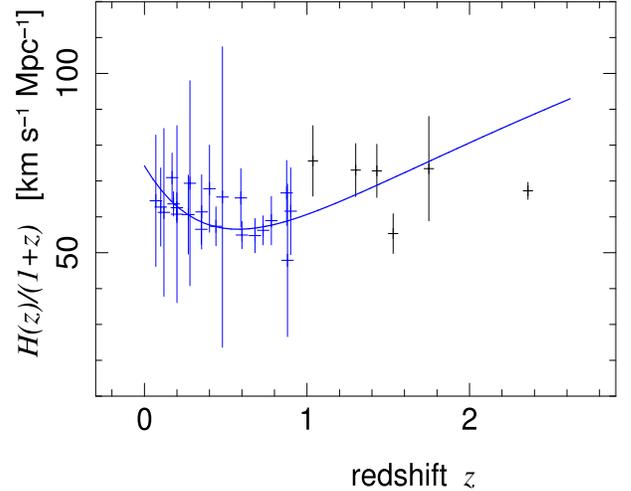}
\caption{\label{fig:lcdm} Expansion rate versus redshift. The crosses are observed data of $\dot{a}=H(z)/(1+z)$ from \cite{simon05}-\cite{rib14}. The blue crosses and line represent data used in fitting Friedmann's Eq.\,(\ref{fri}) of the LCDM model, $\dot{a}=\sqrt{\frac{8\pi G}{3}(\bar{\rho}_{_{m,0}} a^{-1}+\rho_{_\lambda} a^2)+\epsilon}$,  and fitted result, respectively.
}
   \end{center}
\end{figure}

Figure\,\ref{fig:lcdm} shows the observed data of the universe expansion rate $\dot{a}=H(z)/(1+z)$ from independent measurements of the Hubble parameters $H(z)$, where 27 measurements between redshifts $0<z<2$  from \cite{simon05}-\cite{Chuang2012} compiled in \cite{far13}, and $z=2.36$ from \cite{rib14}. It seems that the measured results at redshifts $z<1$ marked by blue crosses in Fig.\,\ref{fig:lcdm} are consistent with the expectation from LCDM: the ongoing accelerating epoch is preceeded by a decelerating phase starting from $z\sim 0.6$, shown by the blue line obtained by fitting Friedmann's equation (\ref{fri}) to the measured $\dot{a}(z)$ at $z<1$. From Fig.\,\ref{fig:lcdm}, however, we also see that there exists a tension between the observed evolution trend of the expansion rates in earlier epoch at redshift $z>1$ and that expected from LCDM. Friedmann's equation (\ref{fri}) predicts that, in the deceleration phase,  the expansion rate should be monotonically increasing with $z$, but in Fig.\,\ref{fig:lcdm} it seems to  turn flat from $z\sim 1$ onwards.

 The measured rate in Fig.\,\ref{fig:lcdm} at $z=2.36$
 \begin{equation}\label{ad23} \dot{a}(z=2.36)=67.3\pm 2.4\hspace{3mm} \mbox{km s}^{-1} \mbox{Mpc}^{-1}
\end{equation}
 is derived by measuring the cross-correlation of quasars with the Lyman $\alpha$ forest absorption, using over 164,000 quasars from the eleventh data release (DR11) of the SDSS-III Baryon Oscillation Spectroscopic Survey (BOSS)  \cite{rib14}. There is another measured rate for high redshift
 \begin{equation}\label{ad234} \dot{a}(z=2.34)=67.1\pm 2.1\hspace{3mm} \mbox{km s}^{-1} \mbox{Mpc}^{-1}\,,
\end{equation}
 which is derived from the same DR11 of SDSS-III BOSS but by a different analysis approach, measuring the flux-correlation of the Ly${\alpha}$ forest using 137,562 quasars \cite{del14}. So far the two  most precise measured rates at $z>2$ are well consistent with each other,  but much lower than the $\simeq 88$\,km s$^{-1}$Mpc$^{-1}$ predicted by LCDM from measurements at $z<1$.
\begin{figure}
\label{fig:ad-s3}
   \begin{center}
\includegraphics[height=6.5cm, width=8cm, angle=0]{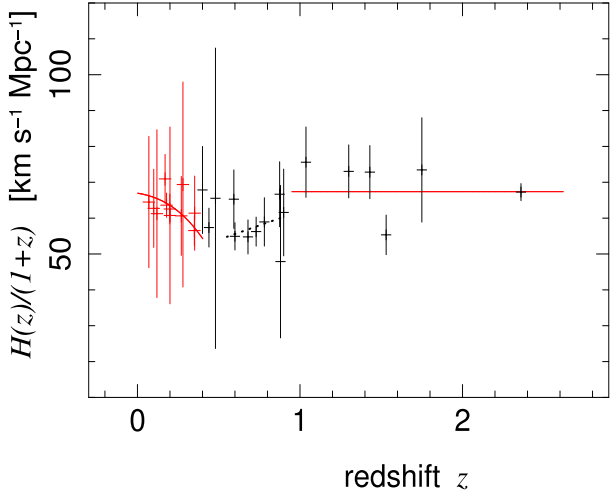}
\caption{\label{fig:ad-s3} Expansion rate versus redshift. The crosses are observed data of $\dot{a}=H(z)/(1+z)$. The red crosses and line at $z<0.4$ represent data used in fitting the acceleration solution $\dot{a}=\sqrt{\epsilon-c_1\exp(-\mu a^2)}$  of the DEMC model and fitted result, respectively. The red horizontal line at $z>1$ presents the constant expansion solution $\dot{a}=\sqrt{\epsilon}$  with $\epsilon$ being estimated from fitting the acceleration solution to the data at $0<z<0.4$. The dotted line is made by polynomial fitting data between redshifts $0.6<z<1$.
}
\end{center}
\end{figure}

 By using the acceleration solution of the DEMC model with undetermined parameters $\epsilon$, $\mu$, and $c_1$
  \[ \dot{a}=\sqrt{\epsilon-c_1\exp(-\mu a^2)}\]
   to fit the measured rates within the ongoing acceleration epoch at $0<z< 0.4$, which are marked by the red crosses in Fig.\,\ref{fig:ad-s3}, we get the predicted critical velocity
 \begin{equation}\label{adc}
 \dot{a}_c=\sqrt{\epsilon}=67.5 \hspace{3mm}\mbox{km s}^{-1}\mbox{Mpc}^{-1}  \end{equation}
  shown by the horizontal red line in Fig.\,\ref{fig:ad-s3}. We can see from Fig.\,\ref{fig:ad-s3} that the varying trend of the observed expansion rate at $z>1$ is well consistent  with the prediction for an equilibrium  state from the expansion equation (\ref{evo}), based on measurements for the accelerating epoch of our universe without using any high-$z$ data. In particular, the predicted $\dot{a}_c=67.5$ km s$^{-1}$Mpc$^{-1}$ is highly coincident with the most precise measured rate $67.3\pm 2.4$\,km s$^{-1}$Mpc$^{-1}$  at $z=2.36$.

  With the fitted Eq.\,(\ref{f1}) from the measured $\dot{a}(z)$ within $0<z<0.4$, we can estimate the current expansion rate (the point on the red line in Fig.\,\ref{fig:ad-s3} at $z=0$) as
\begin{equation}\label{ad0}
\dot{a}_{_0}=H_{_0}=67.0 \hspace{2mm} \mbox{km s}^{-1}\mbox{Mpc}^{-1} \end{equation}
which is well consistent with the observation result from $Planck$+WP \cite{planck13}
%\begin{equation}\label{h00}
\[ H_{_0}=67.3\pm 1.2 \hspace{2mm} \mbox{km s}^{-1}\mbox{Mpc}^{-1}\,, \] %\nonumber \end{equation}
where that estimated from Friedmann's equation (the point on the blue line in Fig.\,\ref{fig:lcdm} at $z=0$)
%\begin{equation}
\[  H_{_0}=74.2 \hspace{2mm} \mbox{km s}^{-1}\mbox{Mpc}^{-1} \] %\end{equation}
is much higher than the $Planck$'s result.

From Eqs.\,(\ref{adc}) and (\ref{ad0}) we get
\begin{equation}\label{ad01}
 \dot{a}_{_0}/\dot{a}_{_c}=0.99 \end{equation}
indicating that the current accelerating universe is close to reaching the next equilibrium period.\\

\section{Cosmological  Parameters}\label{sec:parameter}
From \S2.2 we see that there exist three  physical parameters determining a DEMC universe: the current radius, the constant total rest and mechanical energies ($R_{_0}, E_{rest}, E_{mech}$), or, equivalently, the current radius, current rest energy density, and the constant ratio of mechanical to rest energies ($R_{_0}, \rho_{_0}, \xi)$. To estimate the above parameters we use the following formulae
\begin{eqnarray}
& &\hspace{-70mm}\sqrt{\frac{2\xi}{R_{_0}^2}-c_1\exp[-\frac{3}{2R_0\rho_{_0}(1+z)^2}]} \nonumber\\
\hspace{20mm}=\dot{a}(z)\pm\sigma_{_{\dot{a}(z)}}\hspace{3mm} (0\le z<0.4) \label{fit1}\\
& &\hspace{-70mm}\frac{\sqrt{2\xi}}{R_{_0}}=\dot{a}_c\pm \sigma_{_{\dot{a}_c}} \label{fit2}\\
& &\hspace{-70mm}\sqrt{\frac{8\pi G\rho_{_0}}{3}}=H_{_0}\pm \sigma_{_{H_0}} \label{fit3}
\end{eqnarray}
to fit the data sample of measured $\dot{a}(z)$ and $\sigma_{_{\dot{a}(z)}}$ at $0<z< 0.4$ from \cite{simon05}-\cite{Chuang2012}, $\dot{a}_c=67.1\pm 2.1$  from BOSS/SDSS-III \cite{del14}, and  $H_{_0}=67.3\pm 1.2$  from $Planck$+WP \cite{planck13}. With weighted least-square fitting we get the following  estimations for the fundamental parameters of the universe:  the current radius of universe
\begin{equation}\label{r0}
 \hat{R}_{_0}=(4.1\pm 0.5)\times 10^7 \hspace{3mm} \mbox{Mpc}\,, \end{equation}
current density of rest energy
\begin{equation}\label{rho0} \hat{\rho}_{_0}=(8.5\pm 0.2)\times 10^{-30} \hspace{4mm} \mbox{g cm}^{-3}\,,
 \end{equation}
and  ratio of mechanical to rest energies
\begin{equation}\label{xi1}
 \hat{\xi}=(4.2\pm 1.4)\times 10^7\,. \end{equation}
The standard deviations in the estimations above are obtained with bootstrapped data samples produced by Gaussian sampling from the measured data set.

  The ratio of the current universe radius to the Hubble length $R_{_H}=H_{_0}^{-1}$ can be evaluated as
 \begin{equation}\label{nu}  \nu\equiv R_{_0}/R_{_H}=R_{_0} H_{_0}\,. \end{equation}
 From Eq.\,(\ref{nu}) we see that $\nu$ is also the ratio of the current universe expansion velocity to the speed of light (here $c=1$). Through the process of estimating parameters $(R_{_0}, \rho_{_0}, \xi)$, we can also derive
\begin{equation}\label{velocity}  \hat{\nu}= (9.1\pm 1.2)\times 10^3\,. \end{equation}
That the current radius is much larger than the Hubble length, $\nu \gg 1$, may help to resolve  the tension between the scale of homogeneity in LCDM and the largest dimension $\sim 1240$ Mpc of an observed large quasar group \cite{clo13}.

\begin{figure}
   \begin{center}
\includegraphics[height=5.5cm, width=7cm, angle=0]{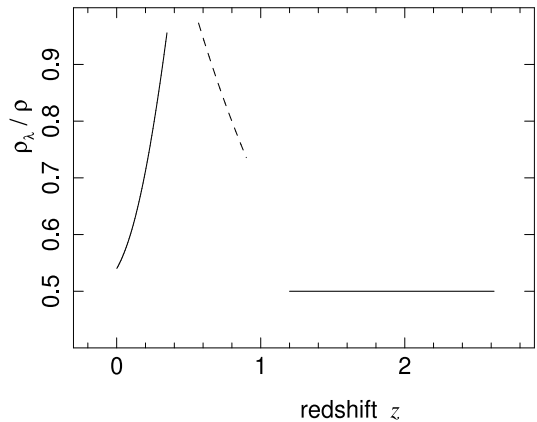}
\caption{\label{fig:rr} Fractional density of dark energy versus redshift.
}
\end{center}
\end{figure}

Between the two expansion solutions shown by the two red lines in Fig.\,\ref{fig:ad-s3}, the equilibrium state at $z>1$ and the accelerating epoch with $z$ less than about 0.6, our universe may go through a phase transition with matter transforming into dark energy.  During a phase transition, the universe expansion  no longer follows the expansion equation (\ref{evo}), but the energy conservation equations (\ref{rest}) and (\ref{f1}) still hold. From Eqs.\,(\ref{rest}), (\ref{f1}) and (\ref{epsilon}) we get
\begin{equation}\label{rr}
\frac{\rho_{_\lambda}}{\rho}=\frac{1}{2}[1-\frac{3(\dot{a}^2-2\xi R_{_0}^{-2})}{8\pi G\rho_{_0}(1+z)}]\,. \end{equation}
With the formula  above, we can evaluate the fractional density of dark energy for all observed region of $z$ from estimated  parameters $\hat{R}_{_0}$,  $\hat{\rho}_{_0}$, $\hat{\xi}$, and  expansion rates $\hat{\dot{a}}$, where $\hat{\dot{a}}(z)$ during the transition period is evaluated from the dotted line shown in Fig.\,\ref{fig:ad-s3} made by polynomial fitting data at $0.6<z<1$. The resulting dark-energy evolution is shown in Fig.\,\ref{fig:rr}: our universe had been in an epoch of equilibrium between gravity and dark energy with the fractional density of dark energy  $\rho_{_\lambda}/\rho\simeq 0.5$ or $\rho_{_d}=\rho_{_m}-\rho_{_\lambda}\simeq 0$ at $z>1$, a phase transition with an increasing fractional density of dark energy  started at $z\sim 1$, and the current acceleration epoch started from $z\sim 0.6$ with a decreasing $\rho_{_\lambda}/\rho$. From Fig.\,\ref{fig:rr} the current fractional density of dark energy is estimated as
\begin{equation}\label{rr0}
\rho_{_\lambda}(z=0)/\rho_{_0}\simeq 0.54 \end{equation}
or
\begin{equation}\label{rd0}
\rho_{_d}(z=0)/\rho_{_0}\simeq 0.08\,. \end{equation}
The current ratio between densities of dark energy and gravity shown in Eq.\,(\ref{rr0}) or Eq.\,(\ref{rd0}), and the ratio of the current expansion rate to the critical expansion rate shown in Eq.\,(\ref{ad01}), both indicate that the current universe is already nearly approaching the next equilibrium epoch.

\section{Motions in the Universe}\label{sec:motion}
\subsection{Cosmological Scale}
In the DEMC model, the global expansion of the universe is jointly  governed by the intrinsic gravitational pull of matter and repulsion from dark energy, i.e. governed by the combined cosmic field $\phi$ determined by the cosmological Eq.\,(\ref{fe})
\[ \nabla^2\phi=4\pi G (\bar{\rho}_{_m}-\rho_{_\lambda}) \]
 which satisfies the global Galileo covariance. From \S\ref{sec:rate} and \S\ref{sec:parameter}, it seems hopeful that the cosmological field equation, Poisson Eq.\,(\ref{fe}), can  properly describe the global motion of a dynamic universe.

\subsection{Local Matter System}
The GR field Eq.\,(\ref{ee0}) proposed by Einstein is to describe the local motion of a free test particle in a gravity field specified by the energy-momentum tensor $T_{\mu\nu}$ of matter, where the universal densities $\bar{\rho}_{_m}$ and $\rho_{_\lambda}$ can be completely ignored, because they are much smaller than the mass  density of a particle and that of bodies constituted by particles.
The motion of an astronomical object within a gravitationally bounded system (a cluster or super cluster of galaxies) relative to the barycenter of the system is determined by the GR field equation (\ref{ee0})
\[ G{_{\mu\nu}}=-8\pi G T_{\mu\nu} \]
which satisfies the local Lorentz invariance.

 To describe the motion of an object in the universe, e.g. the motion relative to the center of another cluster of galaxies, we have to first calculate its motion in the center-of-mass rest coordinates of the local gravitational system by using Eq.\,(\ref{ee0}) with Lorentz invariance, and secondly the universe expansion in the cosmological comoving system by using the cosmological field Eq.\,(\ref{fe}) with Galileo invariance, then superimpose them.

\subsection{Intermediate Case}
The intermediate case that the local gravity specified by the term of $T_{\mu\nu}$ in the field Eq.\,(\ref{ee0}) and the effect of the cosmic field from the source $\bar{\rho}_m-\rho_{_\lambda}$ in the field Eq.\,(\ref{fe}) both have to be considered. A possible intermediate case is a diffuse halo of unparticle dark matter connected with an astronomical object or system.  The density of such a diffuse halo connected gravitationally with an object is  much smaller than the density of the object, but might be much larger than the universal densities $\bar{\rho}_m$ and $\rho_{_\lambda}$. Although the force on a volume of the halo from the combined cosmic field  $\bar{\rho}_m-\rho_{_\lambda}$ is still much weaker than that from the connected object, but it cannot be completely ignored. The halo is pulled by the local gravity of the connected object  but drawn out behind by the cosmic field, and it would be also  slowly expanding driven by the cosmic field but contained by the gravity of the halo itself.
With both Eqs.\,(\ref{ee0}) and (\ref{fe}), we can, in principle, calculate the motion of the halo of dark matter by considering both local gravity and the effect of cosmological pulling, which may help us to  understand the observed deviation between the dark matter halo and the object connected with it.

The most surprising findings in CMB temperature maps observed by {\sl WMAP} and {\sl Planck} missions that challenge the standard model of cosmology are large scale structures alignment with the ecliptic or galactic planes \cite{copi13}. Recently, we find that these abnormal structures can be reproduced from dark matter halos connected with the solar or galactic systems in simultaneously considering the motion of the solar system in the Galaxy or that of the Galaxy in the local cluster of galaxies, and the universe expansion \cite{zhu14}. It is  not necessary to assume an anisotropic primordial universe \cite{planck131}.

 In the early universe, before the formation of structures, the universal matter density $\bar{\rho}_m$ could be comparable with local values $\rho_m$ in $T_{\mu\nu}$, and the state of matter may had effect on the expansion of the universe as considered in the standard model of cosmology, although the content of ordinary matter was only $\sim 5\%$, the universe expansion was still dominated  by the cosmic field.

\section{Discussion}
\subsection{DEMC vs LCDM}
  A dark energy-matter coupling universe described by the field Eq.\,(\ref{fe}) is a well defined mechanical system with constant total and mechanical energies and limited radius, which can be quantitatively evaluated from astronomical measurements. The estimated radius of the current universe in the DEMC model is about ten thousands times larger than the Hubble length, which has the advantage that it resolves the tension between the discovered largest structures and  cosmology (\S\ref{sec:parameter}). The DEMC model provides different kinds of approach to calculate motions for different forms of constituents of the universe:  homogeneous dark energy and gravity, condensated  matter constituted by particles, and condensated unparticle dark matter, and possibly gives a natural interpretation for large scale anomalies in CMB temperature maps (\S\ref{sec:motion}). The evolution of the expansion rate described by the expansion equation (\ref{evo}) in \S\ref{sec:equation} is strictly different from that in LCDM cosmology. As demonstrated in \S\ref{sec:rate}, the DEMC model can fit the observed evolution history of our universe better than the standard model, especially, so far the two high precision measurements for expansion rate at $z=0$ and $z=2.36$ are remarkably coincident with the predictions from the DEMC model.

  LCDM and DEMC are two essentially different models of cosmology with completely different  interpretations for the observed expansion evolution. In LCDM, Einstein's field Eq.\,(\ref{eez}) includes only gravity, one has also to use equations of state  to determine the expansion of the universe with the aid of local pressure of bounded gravitational systems and by the cost of violating the conservation law of energy. In DEMC, on the contrary, no need to confuse the global expansion with local processes, we can interpret the universe expansion with only the cosmological field Eq.\,(\ref{fe}) in Galilean framework because it includes not only gravity but also repulsive dark energy.

  There is hope to finally distinguish  the two kinds of cosmological models in a not too distant future by more high precision  measurements of the expansion rate at the redshift region of $z\ge 1$ (e.g. eBOSS project \cite{ebo}).
  The DEMC scenario, if valid, will radically change our expectations for the origin and future fate of our universe, and affect fundamental physics as well.

\subsection{Origin and Fate}\label{sec:early}
\begin{figure}
   \begin{center}
\includegraphics[height=5cm, width=8.5cm, angle=0]{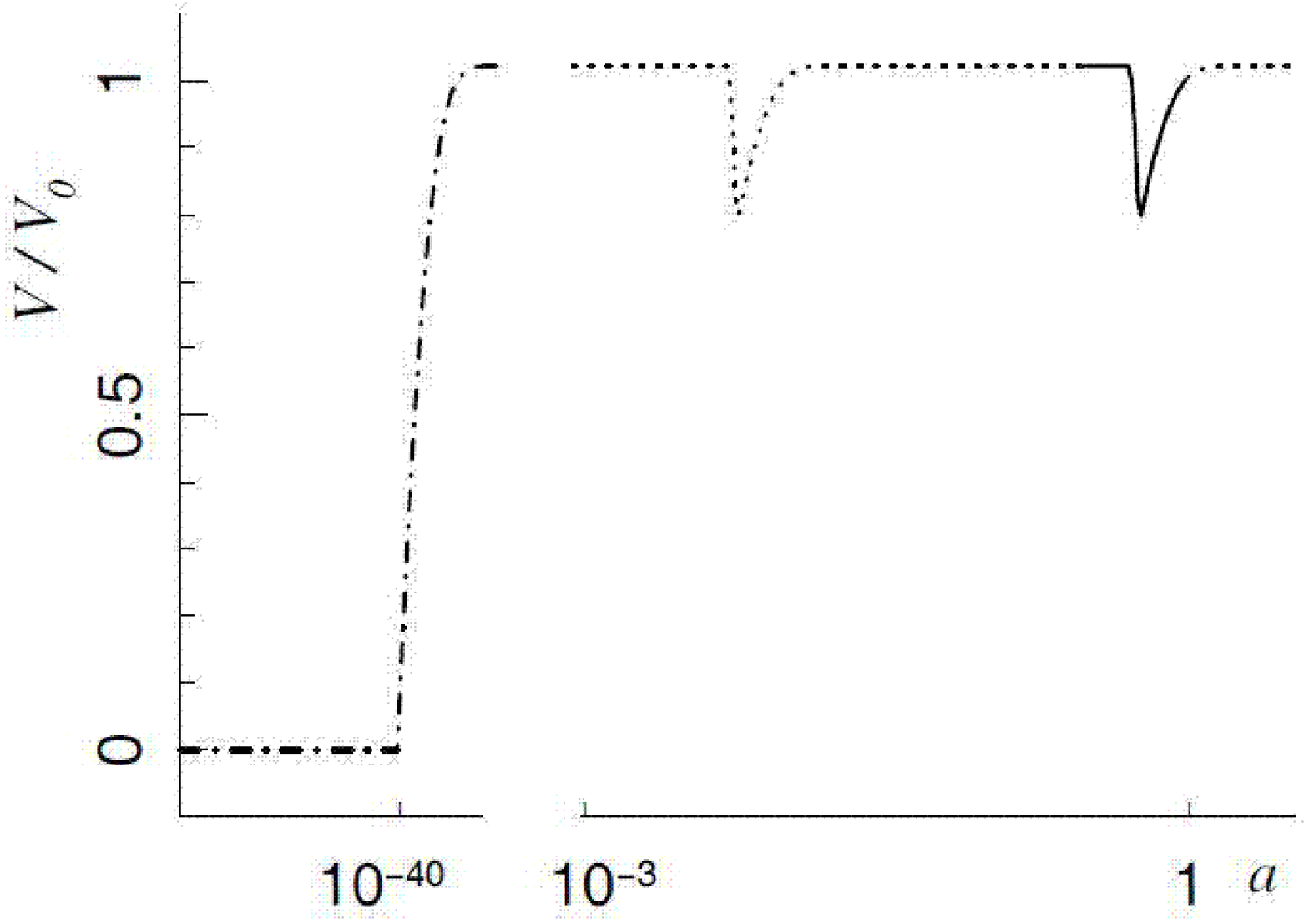}
\caption{\label{fig:evo} A diagrammatic sketch for evolution of a DEMC universe. Scale factor of the universe $a=R/R_{_0}$ where $R$ is radius of the universe, $R_{_0}$ the present radius. The expansion velocity $V=\dot{a} R_{_0}$ and $V_{_0}$ the present velocity.
}
\end{center}
\end{figure}

  Figure\,\ref{fig:evo} schematically illustrates the evolution history of the expansion velocity of a DEMC universe,  where the solid line segment indicates the already observed part of our universe, dotted line segments are expectations from DEMC scenario for earlier and future eras -- normally the universe is constantly expanding but  occasionally interrupted by phase transition. We speculatively put a phase transition at $a\sim 10^{-2}$, whereas the large cold spot with angular size $\sim 10\degree$ \cite{cruz07} and quite a lot  similar cold or hot spots \cite{liu09}  are detected in CMB maps, which may suggest that a cosmic perturbation probably occurred at redshift $z\sim 10^2$.

 For a DEMC universe, inflation can be interpreted as a  phase transition converting rest energy into mechanical energy  occurring in a region of radius $R_{_V}=a_{_V}R_{_0}$, which is a part of the primordial vacuum with energy density $\rho_{_V}$. From rest energy conservation, $\rho_{_V}=\rho_{_0} a_{_V}^{-3}$, we have
 \begin{equation}\label{av} a_{_V}=(\rho_{_V}/\rho_{_0})^{-1/3}\,. \end{equation}
 The vacuum energy density $\rho_{_V}$ predicted by the uncertainty principle sets an upper limit for the energy density $\rho$.  From the so called "the cosmological constant problem" \cite{wei89} we know  that $\rho_{_V}/\rho_{_0}\sim 10^{120}$, consequently $a_{_V}\sim 10^{-40}$ should be a lower limit for the universe dimension, otherwise the energy density of our universe would be larger than the zero-point energy.

 Therefore, our universe could be generated with an initial energy density $\rho_{_V}$ at $z\sim 10^{40}$ by a phase transition in a limited region of $R_{_V}\sim 10^{-40} R_{_0}$, which is placed in a static primordial vacuum consisting of two balanced scalar fields \cite{li11}. Such a scenario for the primordial universe is also help to explain  the observed lack of CMB power on largest scales \cite{liu11,liu13}.

 In the DEMC framework, the primordial universe could be a static vacuum region with radius $R_{_V}$, energy density $\rho_{_V}$, and ratio of mechanical to rest energies $\xi=0$. It is widely accepted that matter (and then inertia) was created during the "reheating" process after inflation \cite{lin90}, then the primordial vacuum constituted by universally balanced attractive and repulsive fields should possess only energy but no inertia. The primordial phase transition broke the equilibrium of the region of $R_{_V}$ in the static primordial vacuum  and resulted in exponential inflation. Through inflation, most of the primordial vacuum energy was transformed into mechanical energy, leading to the current ratio $\xi \gg 1$ as shown by Eq.\,(\ref{xi1}). Except temporary phase transitions, the universe after inflation with "inertial mass" $E_{rest}$ governed by the expansion Eq.\,(\ref{evo}) has been and will be  expanding steadily with a constant rate $\dot{a}_c$.

From Friedmann's Eq.\,(\ref{fri}) and as shown in Fig.\,\ref{fig:lcdm}, a LCDM universe is almost always violently  unstable: in the past, the larger the redshift $z$, the higher the expansion rate ( $\dot{a}\propto \sqrt{z}$ at $z\gg 1$), and in the future, the expansion rate will infinitely increase ( $\dot{a}\propto a$ at $a\gg 1$). In contrast to the standard model, except for transient phase transitions, a DEMC universe is almost always steadily in static equilibrium (pre-inflation vacuum) or stationary expansion with a constant rate (after inflation), that would be more desirable for Einstein.

\subsection{Composite Symmetry}
  Friedmann's equations in the standard model of cosmology  that describe the expansion of the universe, but they are deduced from the GR field equation  that describe the internal motion and structure formation within a gravity system (a dot in the balloon analogy).  That the dark energy cannot be ignored is an essential character in cosmology, thus, as shown in \S\ref{sec:equation}, the general theory of relativity is not suitable to describe the global motion of the universe. It is indeed hard to imagine that the global expansion of universe can be governed by an essentially  local field with Lorentz invariance.  In fact, in constructing the Friedmann-Lemaitre-Robertson-Walker cosmological model from GR, there must be an additional constraint in Einstein field equation, the cosmological principle or  Robertson-Walker metric, to impose a globally uniform space and an universal time upon the universe. Thus, under the cosmological principle, Friedmann's equations can be also derived from Poisson Eq.\,(\ref{ee0a}) and Newton's law of motion \cite{mil34,mil35,mcc34}, or, Newtonian cosmology can correspond to a cosmological theory from GR with Robertson-Walker metric \cite{wei08}.
 Therefore, neither GR nor Newton's theory, but the additional cosmological principle is the real foundation of current models of cosmology.

 In contrast to GR and Newton's theory, homogeneity and isotropy is a natural result of the  DEMC field equation (\ref{fe}) under Galilean framework.
The two coupled and balanced cosmic fields with Poisson Eq.\,(\ref{fe}) provide a simple and sound physical foundation for cosmology with homogeneity, isotropy, energy conservation, non-locality, and non-singularity.

  Therefore, our universe should possess a composite symmetry consisting of global Galileo invariance and local Lorentz invariance.  To describe the motion of an object in the universe, we have to consider both local motion in a gravitationally bound system by Eq.\,(\ref{ee0}) of GR with Lorentz invariance, and the universe expansion in the cosmological comoving system by Poisson Eq.\,(\ref{fe}) with Galileo invariance.  Combining the two kinds of invariances in such calculations, the principle of relativity still holds, or, in other words,  to describe a physical world consisting of different parts with different symmetries, we have to apply different kinds of invariances.

The two kinds of cosmological models, LCDM and DEMC, have a similar thermal and nucleosynthesis histories after inflation, but completely different from each other before inflation.  If the DEMC scenario is valid,  there would be no hot Big Bang for our universe as shown in Fig.\,\ref{fig:evo}, and  the Planck era may never exist during the whole history of our universe. Consequently, instead of seeking for a grand unification, we should accept a composite scheme for our universe: the cosmic-, macro-, and micro- worlds are governed by Galileo, Lorentz, and Yang-Mills invariances, respectively.\\

\noindent{\large\bf Acknowledgments} This work is supported by the National Natural Science Foundation of China (Grant No. 11033003). Profs. Mei Wu and Charling Tao are thanked for helpful comments and suggestions.

    \vspace{3mm}

     \end{document}